\documentclass{article}

\usepackage[english]{babel}
\usepackage{amsmath}
\usepackage{graphicx}
\usepackage[colorinlistoftodos]{todonotes}
\usepackage[colorlinks=true, allcolors=blue]{hyperref}
\usepackage{amsthm,amsfonts,amscd,amssymb,eucal,latexsym,mathrsfs}
\usepackage{epstopdf}
\usepackage{tikz}
\usepackage[utf8]{inputenc}
\usepackage{pdfpages}
\usetikzlibrary{calc}
\usepackage{caption}
\usepackage{subcaption}
\usepackage{authblk}

\title{The Dynamics of  Interacting Swarms \footnote{NRL/MR/6790--18-9782}}

\author[2]{Carl Kolon}
\author[1]{Ira B. Schwartz}
\affil[1]{US Naval Research Laboratory\\
Plasma Physics Division\\ Nonlinear Systems Dynamics Section\\
Code 6792\\
Washington, DC 20375\\
Contact: ira.schwartz@nrl.navy.mil}
\medskip
\affil[2]{United State Naval Academy\\
Department of Mathematics\\
Chauvenet Hall \\
572C Holloway Road\\
Annapolis, MD 21402-5002}

\date{}

\begin{document}
\maketitle
\begin{abstract}
Swarms are self-organized dynamical coupled agents which evolve from simple
rules of communication. They are ubiquitous in nature, and becoming more
prominent in defense applications. Here we report on a preliminary study of
swarm collisions for a  swarm model in which each agent is self-propelling but
globally communicates with other agents. We generalize previous models by
investigating the interacting dynamics  when delay
is introduced to the communicating agents.  One of our major  findings is that
interacting swarms are far less likely to flock cohesively if they are coupled
with delay. In addition, parameter ranges based on coupling strength,
incidence angle of collision, and delay change dramatically for other swarm interactions
which result in flocking, milling, and scattering.  
\end{abstract}

\tableofcontents

\newpage
\section{Introduction}
The emergence of complex dynamical behaviors from simple local interaction rules
between pairs of agents in a group is a widespread phenomenon over a range of
application domains, and results in coherent behavior called a swarm. Many
striking examples can be found in biological 
systems, from the microscopic (ex., aggregates of bacterial cells or the
collective motion of skin cells in wound healing)
\cite{Budrene1995,Polezhaev2006,Lee2013} to large-scale aggregates of fish,
birds, and even humans \cite{Tunstrom2013,Helbing1995,Lee2006}.  These systems
are particularly interesting to the robotics community because they allow
simple individual agents to achieve complex tasks in ways that are scalable,
extensible, and robust to failures of individual agents. In addition, these
aggregate behaviors are able to form and persist in spite of complicating
factors such as communication delay and restrictions on the number of
neighbors each agent is able to interact with, heterogeneity in agent
dynamics, and environmental noise. These factors define how  swarm
behaviors are effected by changes due to internal and external actors, allow
the swarm to change its patterns and
function~\cite{szwaykowska2015patterned,szwaykowska2015collective,szwaykowska2016collective,hindes2016hybrid},
as well as break apart and reform
~\cite{morgan2005dynamic}.

A number of studies show that even with simple interaction protocols, swarms
of agents are able to converge to organized, coherent behaviors. Existing
literature on the subject provides a wide selection of both agent-based
\cite{Helbing1995,Lee2006,Vicsek2006,Tunstrom2013} and continuum models
\cite{Edelstein-Keshet1998,Topaz2004,Polezhaev2006}. One of the earliest
agent-based models of swarming is Reynolds's \textit{boids}
\cite{Reynolds1987}, which simulates the motion of a group of flocking
birds. The boids follow three simple rules: collision avoidance, alignment
with neighbors, and attraction to neighbors. Since the publication of
Reynolds's paper, many models based on ``zones'' of attraction, repulsion,
and/or alignment have been used as a means of realistically modeling swarming
behaviors \cite{Miller2012,Tarras2013,Viragh2014}. Systematic numerical
studies of discrete flocking based on alignment with nearest neighbors we
carried out by Vicsek \textit{et al.}  \cite{Vicsek1995}. Stochastic
interactions between agents are modeled in \cite{Nilsen2013}. In recent years,
improved computer vision algorithms have allowed researchers to record and
analyze the motions of individual agents in biological flocks, and formulating
more accurate, empirical models for collective motion strategies of flocking
species including birds and fish \cite{Ballerini2008,Katz2011,Calovi2014}.

Despite the multitude of available models, how group motion properties emerge
from individual agent behaviors is still an active area of research. For
example, \cite{Viscido2005} presents a simulation-based analysis of the
different kinds of motion in a fish-schooling model; the authors map phase
transitions between different aggregate behaviors as a function of group size
and maximum number of neighbors that influence the motion of each ``fish''. In
\cite{Lee2006}, the authors use simulation to study transitions in aggregate
motions of prey in response to a predator attack.

Interaction delay is a ubiquitous problem in both naturally-occurring and artificial systems, including blood cell
production and coordinated flight of bats \cite{Martin2001,Bernard2004,Monk2003,Giuggioli2015,Forgoston2008}.
Communication delay can cause emergence of new collective motion patterns and lead to noise-induced switching between bistable
patterns \cite{Romero2011,Romero2012,Lindley2013a}; this, in turn, can lead to instability in robotic swarming systems
\cite{Viragh2014,Liu2003}. Understanding the effects of delay and latency is
key to understanding many swarm behaviors and functions in natural,
as well as engineered, systems \cite{szwaykowska2016collective}. As another
controlling factor of swarm function,
network communication topology maybe used to improve and generalize swarm
function. For example in~\cite{hindes2016hybrid}, it was shown how to take
known basic swarm modes, such as flocking, milling and rotating states, and
use network topology to create new hybrid states that form new functions,
such as shielding important swarm controlling agents.

In addition, many models make the mathematically simple but physically implausible assumption that swarms are
globally coupled (that is, each agent is influenced by the motion of all other agents in the swarm)
\cite{Motsch2011,Chen2011,Chen2014,Lee2006,Vecil2013a}. Global coupling is easier to analyze and a
reasonable assumption in cases of high-bandwidth communication, with a sufficiently small number of
agents. In contrast, we are interested in the collective motion patterns that emerge when global
communication cannot be achieved. New behaviors can unexpectedly emerge when the communication structure
of a network is altered, as in \cite{VonBrecht2013}, where the stability of solutions for compromise dynamics over an
Erd{\"o}s-Renyi communication network is considered. However, in our system, we show robustness of emergent motion
patterns to loss of communication links in presence of delayed coupling~\cite{szwaykowska2016collective}.

A third effect, which we do not consider here, is agent heterogeneity. Most existing work assumes
that the members of the swarm are identical. However, many practical
applications involve swarms that are composed of agents with differing
dynamical properties from the onset, or that become different over time due to
malfunction or aging. Swarm heterogeneity leads to interesting new collective
dynamics such as spontaneous segregation of the various populations within the
swarm; it also has the potential to erode swarm cohesion. In biology, for
example, it has been shown that sorting behavior of different cell types
during the development of an organism can be achieved simply by introducing
heterogeneity in inter-cell adhesion properties
\cite{Steinberg1963,Graner1993}. In robotic systems, allowing for
heterogeneity in dynamical behaviors of swarm agents gives greater flexibility
in system design, and is therefore desirable not only from a theoretical but
also from a practical point of view~\cite{hindes2016hybrid}. 

In this  preliminary work we are interested in swarms that are  interacting. That is, the entire
swarm is split into two parts initially, and then the two components approach each other as flocks. We
then are interested in understanding the results based on physical parameters,
such as attraction/repulsion length scales, delayed communication, and
coupling strength. The model for the swarm we use is based on the the employed
in~\cite{DOrsogna:2006}, which  describe a mathematically  swarm model using
the Morse potential. Recently, the authors in~\cite{Armbruster:2017} studied
the effects of swarm collisions using the Morse potential as a function of
incident angle of the interacting flocks. However, in any robotic swarm there
must be some delay between sensing other agents,  and control actuation. We
investigated the result of applying delay to the coupling terms of a swarm
model. As a result, every agent behaves as though the other agents are at the
positions they were at $\tau$ time units before. We found that the flocking
state is unlikely following collisions of delay-coupled swarms, even for
fairly small values of $\tau$. The implication is that it should be possible
to have one swarm modify the intent of another by either capturing the swarm,
or redirecting the mean direction of the flock.


\section{The Basic Swarm Model}
The swarm model we used for this preliminary study  follows the model in
\cite{DOrsogna:2006}. It  is a system of $N$ agents in $n$-dimensional space
with position vectors $x_i\in \mathbb{R}^{n} $, acting under the following equation of motion:
\begin{equation}\label{instantaneous}
\ddot{x}_i=(\alpha-\beta\lvert\dot{x}_i\rvert^2)\dot{x}_i-\frac\lambda N\sum_{j=1,i\neq j}^N \nabla_{x_i}U(x_i,x_j),
\end{equation}

\noindent where $\alpha, \beta, \lambda$ are constants,
$U:\mathbb{R}^{2n}\rightarrow\mathbb{R}$ is a potential function of the two
agents' position, and $\nabla_{x_i}$ represents taking the gradient with
respect to $x_i$. Many potential functions can be used, as in \cite{Mogilner2003}, but a useful choice is the scaled Morse potential:
\begin{equation}\label{potentialeqn}
U(x_i,x_j)=C\exp(-\lvert x_i-x_j\rvert/l)-\exp(-\lvert x_i-x_j\rvert).
\end{equation}
In particular, the parameter $l$ can be used to set a ratio of repulsive and
attractive length scales.

\begin{figure}[h]
\centering
\begin{subfigure}{.4\textwidth}
\includegraphics[width=\textwidth]{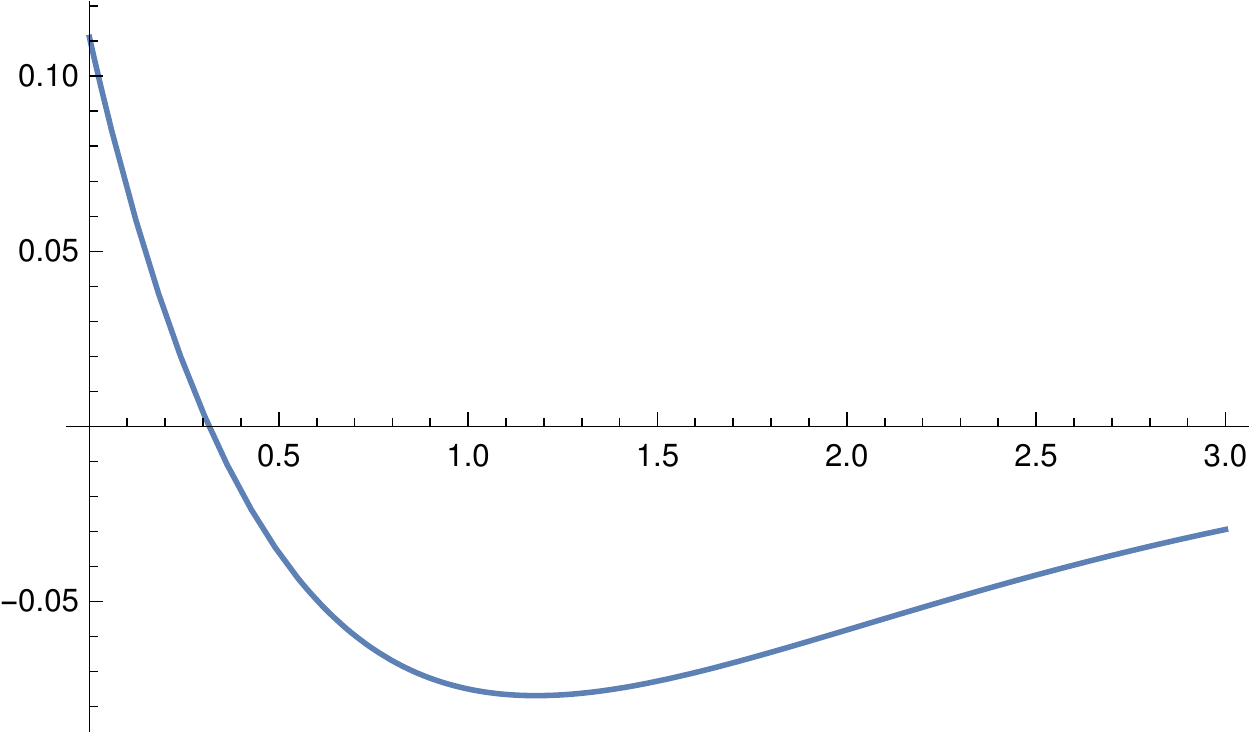}
\caption{Morse potential as a function of distance.}
\end{subfigure}\hfill
\begin{subfigure}{.4\textwidth}
\includegraphics[width=\textwidth]{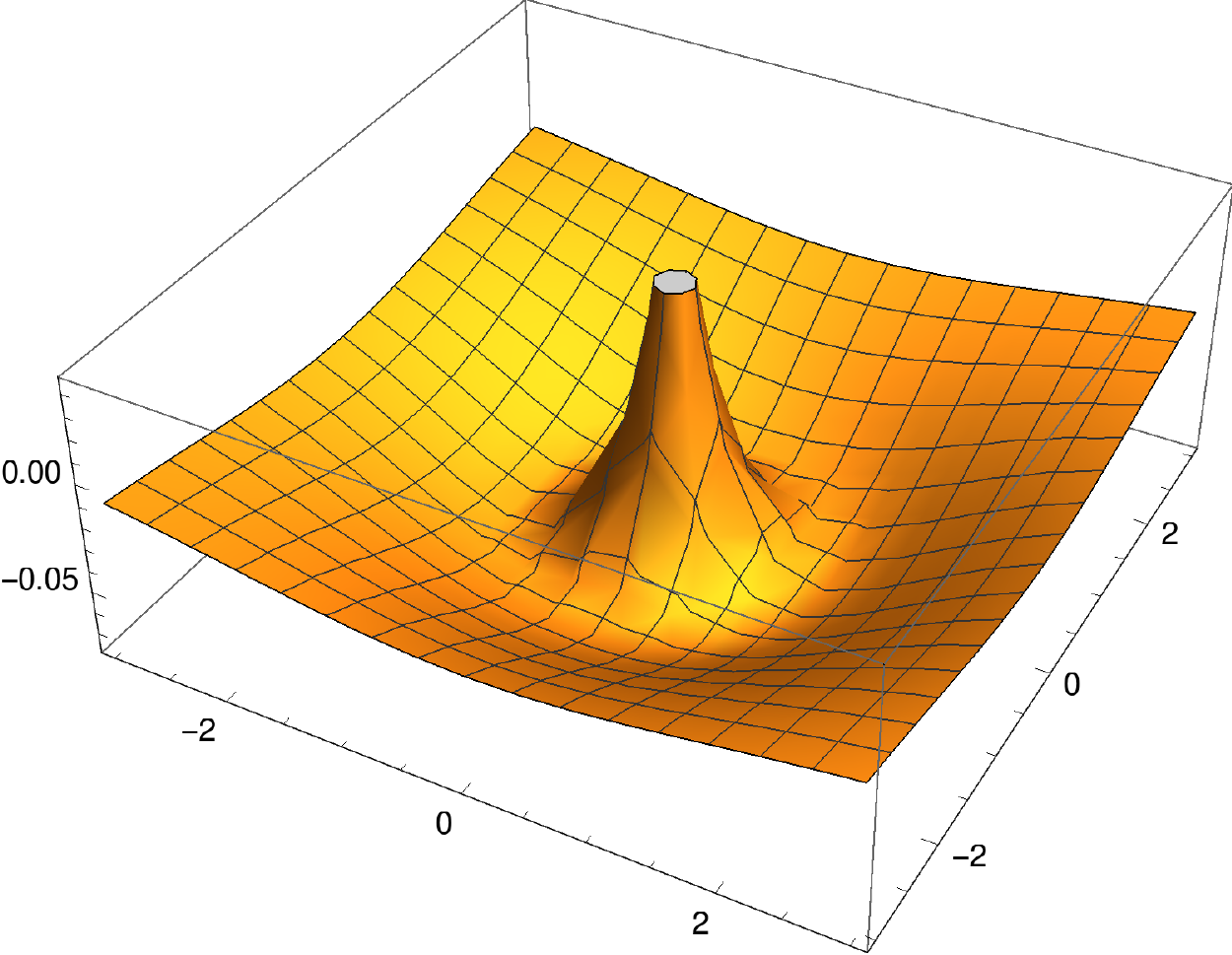}
\caption{Morse potential of a variable point on a plane and a point at the origin.}
\end{subfigure}
\caption{\label{morse}The Morse potential with $C=\frac{10}9$ and $l=\frac34$.}
\end{figure}

Following~\cite{Armbruster:2017}, we fix the values of $\alpha$, $\beta$, $C$,
and $l$. We also constrain our swarm to the plane. We choose $\alpha=1$,
$\beta=5$, $C=\frac{10}9$, $l=\frac34$. This means that, at long ranges,
$\nabla_{x_i}U(x_i,x_j)$ will be attractive, and at short ranges it will be
repulsive (see Fig.~\ref{morse}). Furthermore, the parameters choice places the swarm
in the region which  is defined as "catastrophic"~\cite{DOrsogna:2006}, meaning that
increasing the number of agents decreases the space between agents, so that
the total size of the swarm approaches a limit in space. 

When $\nabla_{x_i}U(x_i,x_j) \equiv 0$ for all $i,j\leq N$ with $i\neq j$, the
dynamics is based solely on self-propulsion, and the resulting configuration
of agents is called a flock \cite{Levine:2000}. That is, all of the agents
when started in parallel, will translate with the same asymptotic velocity and
direction. In the attraction-repulsion model, a flock undergoes translating
motion at a constant speed $\sqrt{\alpha/\beta}$, and can withstand
sufficiently small perturbations \cite{Carrillo:2014}. However, due to
translation symmetry, there will always be one eigenvalue in the linearization
about the flocking state equal to zero, which means the stability is neutral
at best. Hence small perturbations may destabilize the flocking dynamics \cite{mier2012coherent}.

Since we are interested in flock collisions, two separate flocks are initialized and pointed towards each other with some incident angle, $\theta$. We are interested in the behavior after collision, and how it varies for varying $\lambda$, $\theta$, and $\tau$.

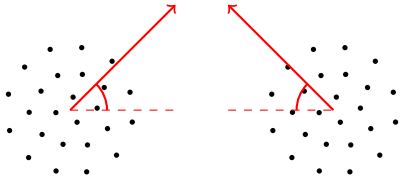
\begin{figure}[h]
\centering
\begin{tikzpicture}[scale=.7]
\foreach \position in {(0.5122,0.0375198),(1.17914,-0.226873),(0.647069,0.430066),(-1.17235,0.302789),(1.13673,0.340089),(-0.778274,-0.0416182),(-0.236511,0.657345),(0.0548554,0.246795),(-1.15108,-0.388026),(0.87992,-0.855647),(-0.534164,-0.462043),(-0.271585,-0.0473607),(0.31269,-1.17263),(0.21905,1.18738),(-0.865201,0.818041),(-0.553714,0.361024),(-0.789311,-0.901525),(0.32852,-0.661597),(-0.268163,-1.15609),(0.708726,-0.351244),(-0.135683,-0.648939),(-0.380444,1.15435),(0.231453,0.679006),(0.795655,0.928953),(0.130461,-0.229757),(5.5122,0.0375198),(6.17914,-0.226873),(5.64707,0.430066),(3.82765,0.302789),(6.13673,0.340089),(4.22173,-0.0416182),(4.76349,0.657345),(5.05486,0.246795),(3.84892,-0.388026),(5.87992,-0.855647),(4.46584,-0.462043),(4.72842,-0.0473607),(5.31269,-1.17263),(5.21905,1.18738),(4.1348,0.818041),(4.44629,0.361024),(4.21069,-0.901525),(5.32852,-0.661597),(4.73184,-1.15609),(5.70873,-0.351244),(4.86432,-0.648939),(4.61956,1.15435),(5.23145,0.679006),(5.79566,0.928953),(5.13046,-0.229757)}
	\fill \position circle (0.05cm);
	\draw[red,->,thick] (0,0) -- (2,2);
	\draw[red,->,thick] (5,0) -- (3,2);
	\draw[red,thick] (.7,0) arc (0:45:.7);
	\draw[red,dashed] (0,0) -- (2,0);
	\draw[red,thick] (4.3,0) arc (180:135:.7);
	\draw[red,dashed] (3,0) -- (5,0);
\end{tikzpicture}
\caption{\label{varytheta}The incident angle of the collision, $\theta$, which we vary as a parameter.}
\end{figure}

\begin{figure}[h]
\centering
\begin{subfigure}{.3\textwidth}
\includegraphics[width=\textwidth]{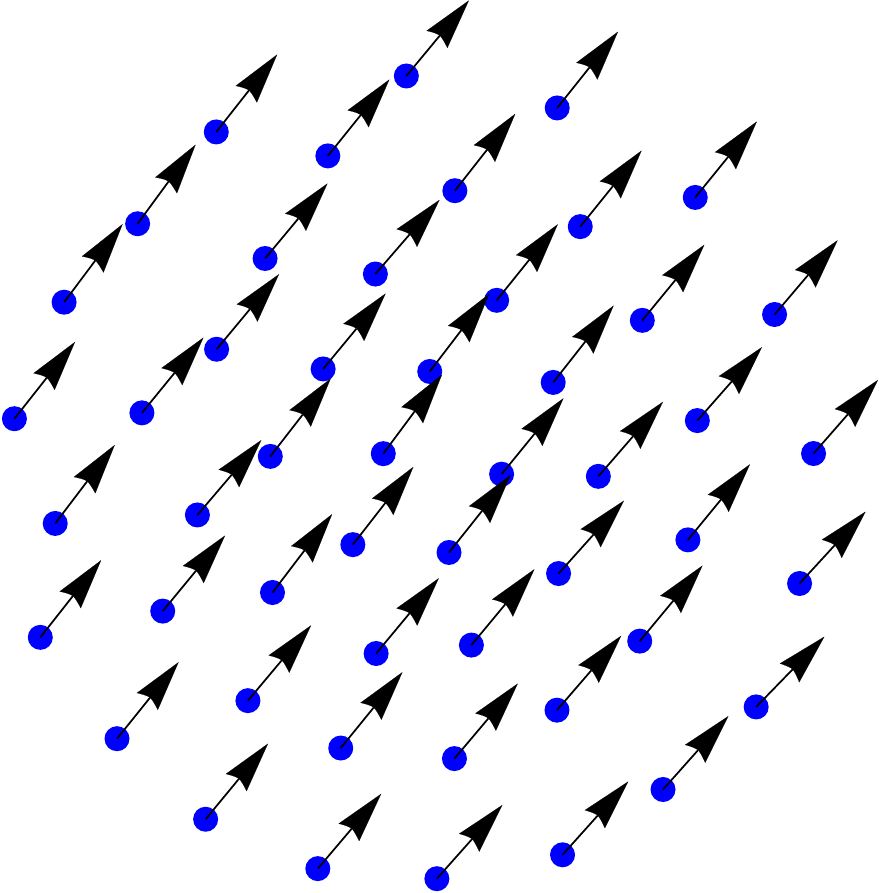}
\caption{Flocking.}
\end{subfigure}\hfill
\begin{subfigure}{.3\textwidth}
\includegraphics[width=\textwidth]{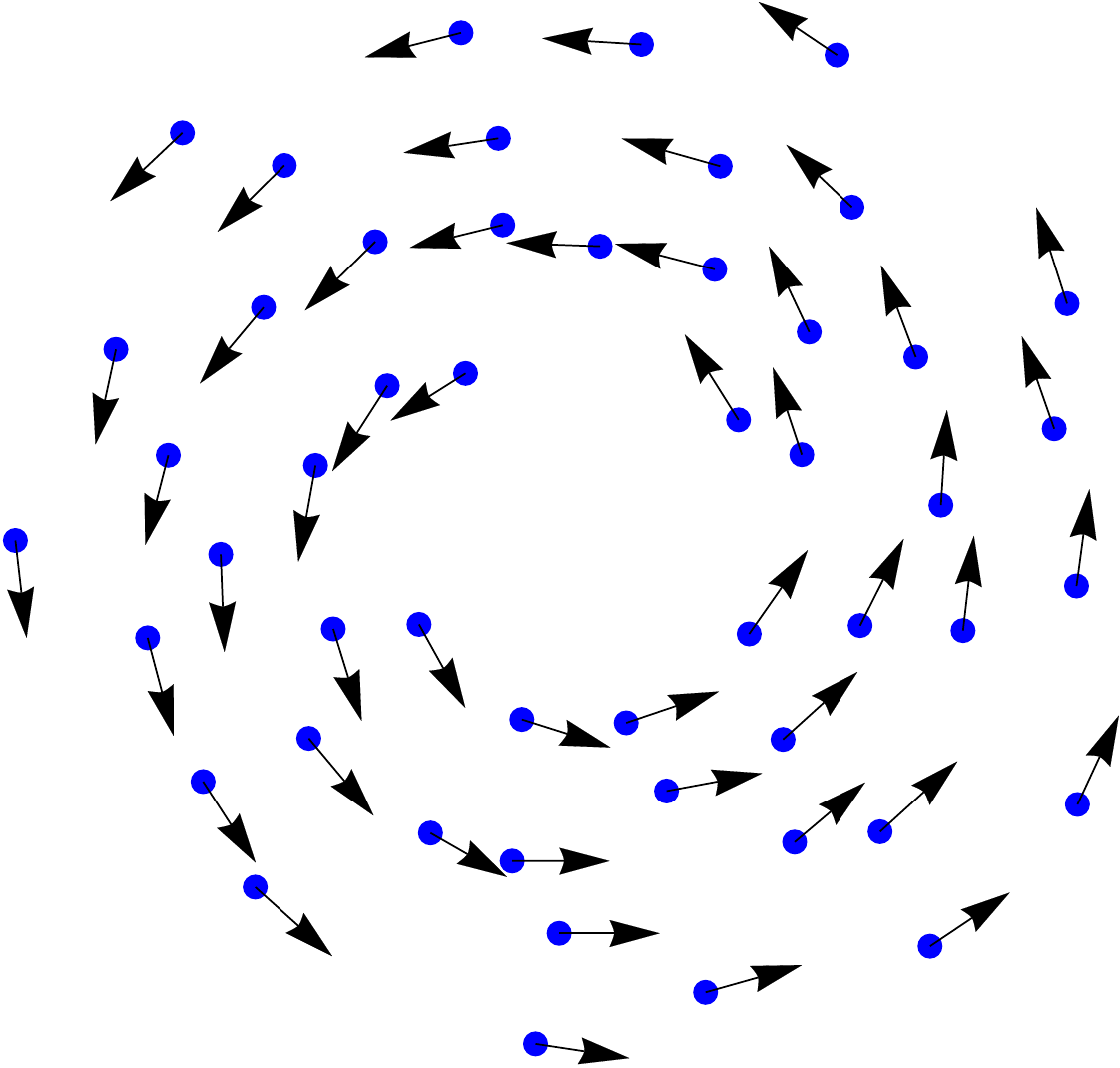}
\caption{Milling.}
\end{subfigure}
\begin{subfigure}{.3\textwidth}
\includegraphics[width=\textwidth]{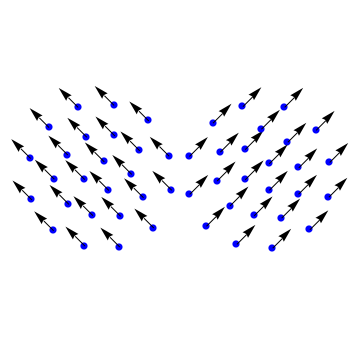}
\caption{Scattering.}
\end{subfigure}
\caption{\label{flockmill}Flocking, milling, and scattering behavior in a Morse swarm.}
\end{figure}

Flocking is not the only behavior that swarms display. They also can perform
milling behavior, where the agents rotate around a stationary center of
mass. To differentiate between the two behaviors, we calculate the
polarization of the swarm as the directed sum of velocities. Specifically, for
a system of $N$ agents with position vectors $x_i$, the polarization $P$ of
the flock is defined as:
\[P(x_1,x_2,\ldots,x_N)=\left\lvert\frac{\sum_{i=1}^N\dot{x}_i}{\sum_{i=1}^N\lvert\dot{x}_i\rvert}\right\rvert\]
Two limiting behaviors can be immediately seen to be reflected in the
polarization. If the velocities are coherent, we expect a value of $P \approx
1$. If the velocities point in opposite directions, we expect $P\approx
0$. Therefore, polarization is a good measure of whether the system displays
flocking ($P\approx1$), or milling ($P\approx0$). Additionally, two colliding
swarms can scatter, in which they do not form a coherent group which either
flocks or mills. In this case, we expect $P$ to be neither close to 0 nor
close to 1. See Fig.~\ref{flockmill} for examples of the three types of swarm interactions.

\section{Delay-Coupled Swarms}
As mentioned above, since robotic systems possess latency in both
communication and control actuation, we generalized the model to include a time delay $\tau$ in the interaction term, creating the following delay differential equation from equation (\ref{instantaneous}):

\begin{equation}\label{delaycoupled}\ddot{x}_i(t)=(\alpha-\beta\lvert\dot{x}_i(t)\rvert^2)\dot{x}_i(t)-\frac\lambda N\sum_{j-1,i\neq j}^N \nabla_{x_i}U(x_i(t),x_j(t-\tau)),\end{equation}
with $U$ defined in Eq.~(\ref{potentialeqn}). Since this is a delay
differential equation, it requires an initial  history function of position
and velocity defined over
$[-\tau,0]$, for which we used translating motion of the two flocks along
linear trajectories. We simulated two flocks of 25 agents, each swarm  on a
collision path with the other. We varied the values of $\lambda$, the coupling strength, and $\theta$, the incident angle of the collision.

\begin{figure}[h]
\centering
\begin{tikzpicture}[scale=2]
    \node[anchor=south west,inner sep=0] at (0,0) {\includegraphics[width=8.64cm]{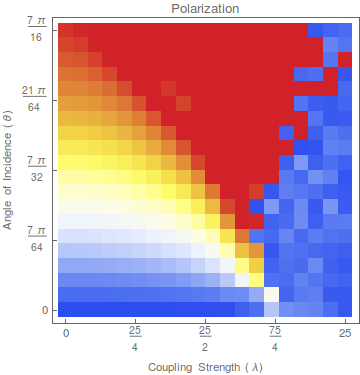}};
    \node[anchor=south west,inner sep=0] at (4.5,.8) {\includegraphics[width=1cm]{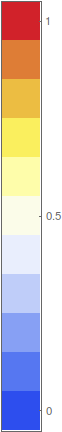}};
    \draw (3.28,.7) .. controls (2.65,2.45) .. (.7,4.2);
    \draw (3.28,.7) .. controls (3,2) .. (4.23,3.6);
    \draw (3.28,.7) .. controls (2.65,2.45) and (3.4,2.4).. (3.7,4.24);
    \node at (2.7,3.5) {Flocking};
    \node at (1.5,2) {Scattering};
    \node at (3.7,2) {Milling};
    \node[rotate = 60] at (3.7,3.3) {Metastability};
\end{tikzpicture}
\caption{\label{0delay}Polarization as a function of $\lambda$ and $\theta$. Note the distinct regions of flocking, milling, scattering, and metastability.}
\end{figure}
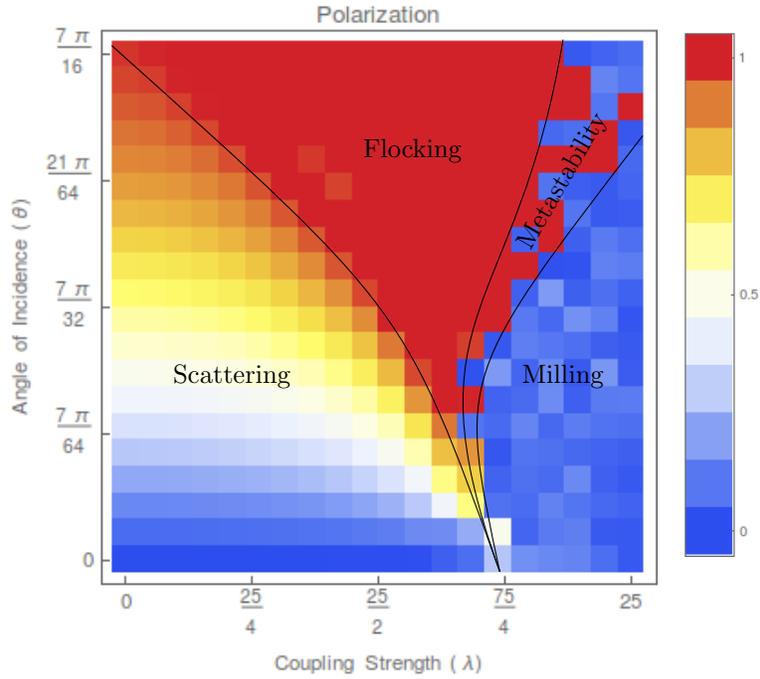

As a benchmark, we first simulated the results  by
performing the experiment without delay;i.e., $\tau=0$, corresponding to a
system with instantaneous interactions. We simulated 400 collisions and
recorded the polarization after  transients were removed, about 100 time
units. We generated the plot in Fig.~\ref{0delay} showing how polarization
can be used to quantify the resulting dynamics after the swarms collide.

From Fig.~\ref{0delay}, we immediately observe several large scale  features
in parameter space. For low values of $\lambda$, scattering
is likely unless the swarm motion is almost parallel in its initial state. For
intermediate values of $\lambda$, flocking is likely to be observed, while for high values of
$\lambda$, milling dominates the parameter region. A region of metastability exists between the
flocking and milling states. The black lines are used to guide the eye to show
the boundaries separating the final swarm states.

\begin{figure}[h!]
\centering
\includegraphics[width=\textwidth]{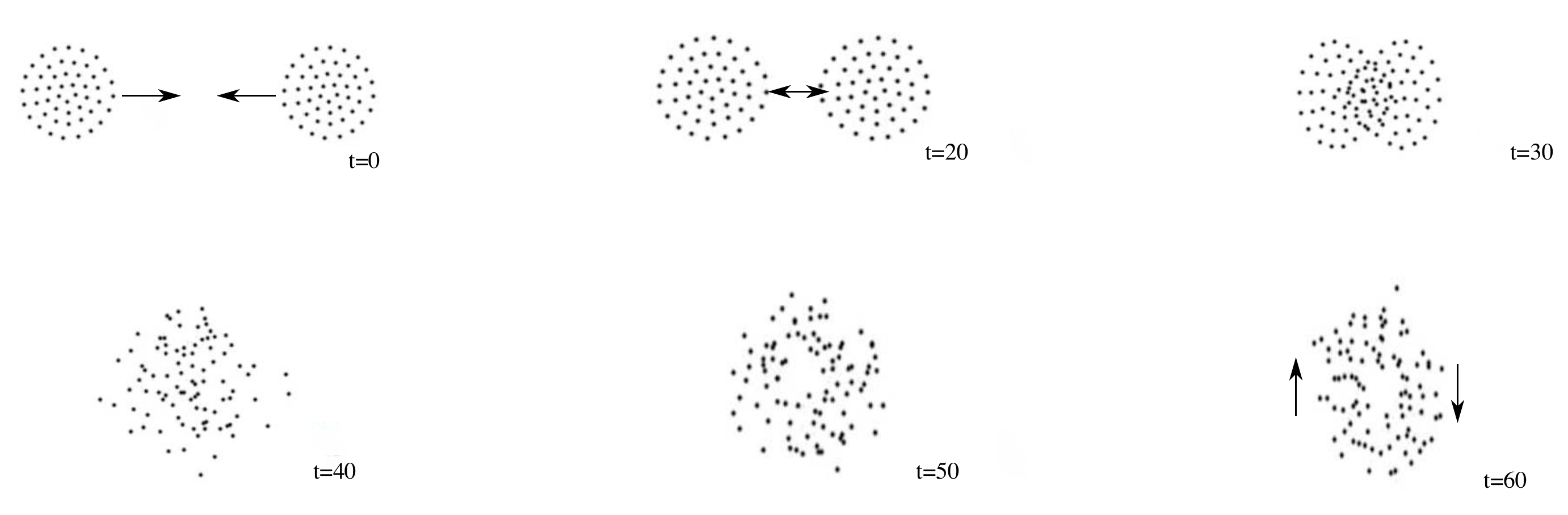}
\caption{Time snapshots of swarm capture to a milling state. Delay here is
  $\tau=0.1$, and other parameters were chosen in the milling regime of Fig.}
\label{fig:comp_milling}
\end{figure}

To see how the milling comes into play for the delay-coupled swarm, we start
the flocks approaching each other at zero incidence
angle. Figure~\ref{fig:comp_milling} illustrates the progression from
the initial flocks to the final milling swarm. It shows the potential of how
one might use a defensive swarm to stop another by converting its flocking
state, which is a traveling center of mass, to a milling state with a center
of mass that is stationary.

\begin{figure}[t]
\centering
\begin{subfigure}{.425\textwidth}
\includegraphics[width=\textwidth]{polarization500s}
\caption{$\tau=0$}
\end{subfigure}
\begin{subfigure}{.425\textwidth}
\includegraphics[width=\textwidth]{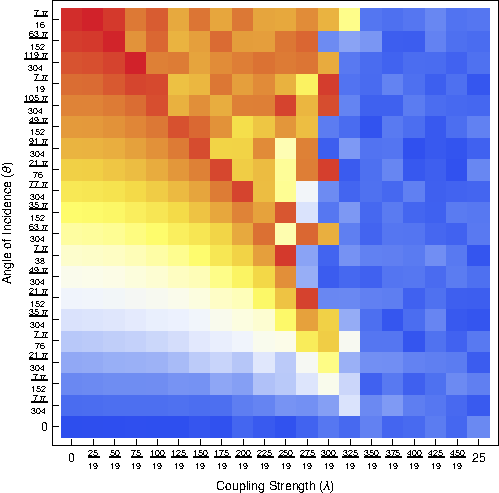}
\caption{$\tau=0.1$}
\end{subfigure}\\
\begin{subfigure}{.425\textwidth}
\includegraphics[width=\textwidth]{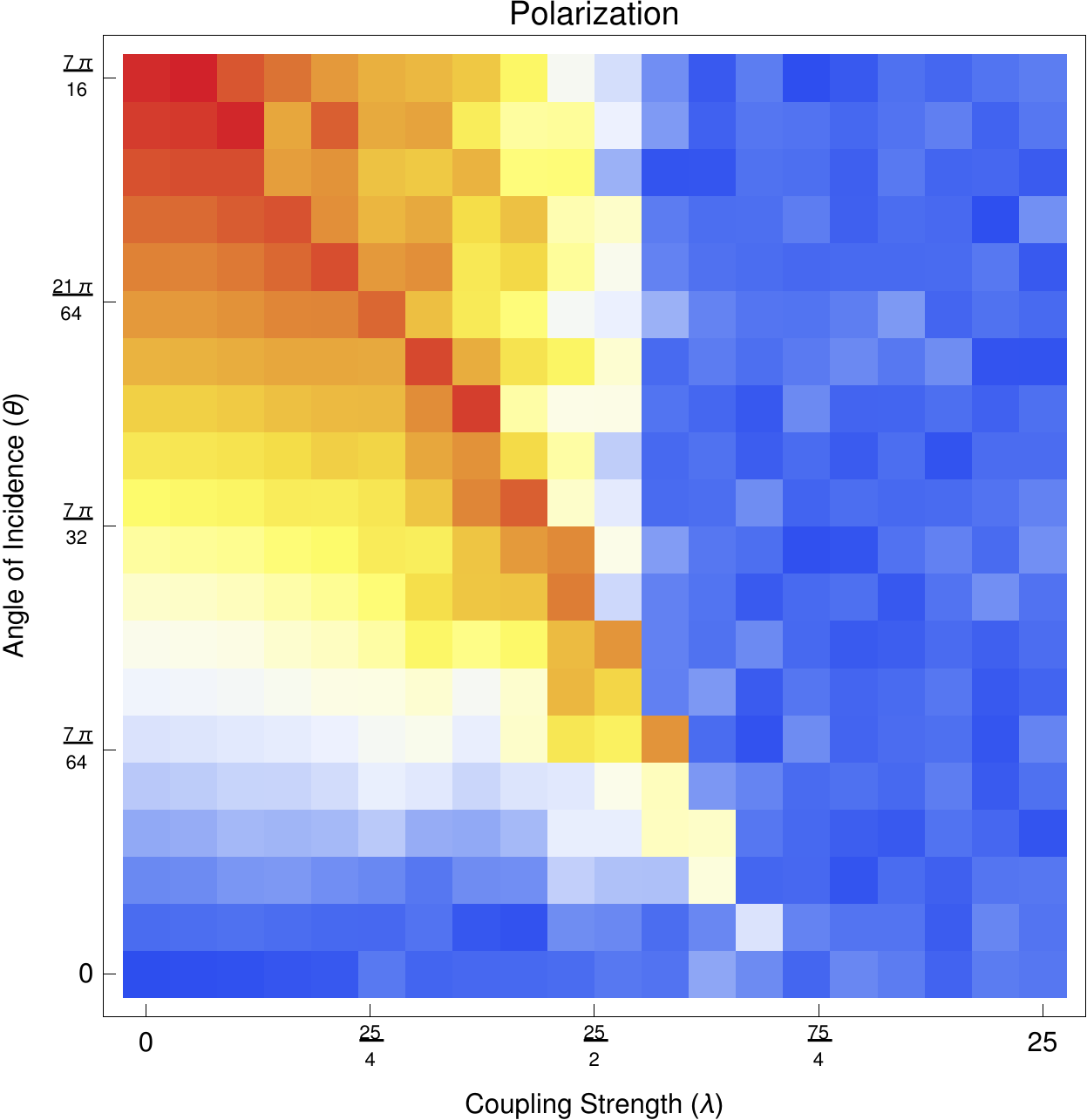}
\caption{$\tau=0.2$}
\end{subfigure}
\caption{\label{tauplots}Polarization as a function of $\lambda$ and $\theta$ for 3 distinct values of $\tau$.}
\end{figure}

In exploring the effects of delay on the parameter basins of swarm final states, we computed the polarization for  three distinct values of delay:
$\tau=$ 0, 0.1, and 0.2. The results of  the average polarization plots using the
above procedure are in Fig.~\ref{tauplots}.

We can note several things about the introduction of delay in panels of
Fig.~\ref{tauplots}. Reference from the $\tau=0$ case, one see a gradually
shrinking of the flocking region, with an increase in milling parameters. Even
at the relatively small delay of $\tau=0.2$, flocking behavior nearly vanishes. Instead, scattering behavior becomes more likely for low $\lambda$, and milling behavior becomes more likely for high $\lambda$. 

\begin{figure}
\centering
\includegraphics[width=.425\textwidth]{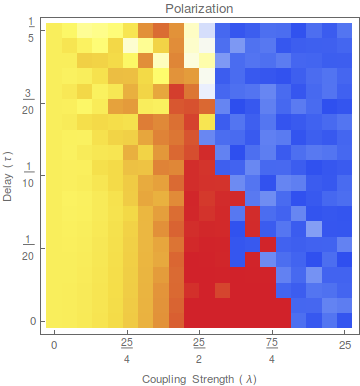}
\caption{\label{lambdatau}A plot of polarization as a function of $\lambda$ and $\tau$ for fixed $\theta$. Note that flocking behavior all but disappears above $\tau\approx0.1$.}
\end{figure}

To check the overall dependence of swarm final state on both coupling strength
and delay,  we fixed the incidence angle $\theta$ at $\frac\pi4$, and varied
$\tau$ finely, to observe polarization as a function of $\tau$ and
$\lambda$. The results are shown in Fig.~\ref{lambdatau}. This confirms that as delay increases steadily, milling behavior largely replaces flocking behavior.

\section{Conclusion}
The preliminary results of this work reveal certain features about how
interacting swarms upon collision modify their behavior.
One major finding is that the existence of a swarming flocking state  depends
strongly on the delayed interaction between agents. If  delays can be induced
into the communication network of a flocking swarm, the results show that even
small delay disrupts the motion of the center of mass, thus disabling the
flocking state.

In the case of globally coupled  interacting swarms which are equal in number,  the resulting swarm states may be in one of
three states: Flocking, Milling, and Scattered. However, the results from this
study depend heavily on the initial states, such as velocity and incidence
angle, in addition to delay magnitude. Nonetheless, the final states of
colliding swarms may result in modifying the center of mass velocity,
change the navigation course of the swarm from its original path, and capture
another swarm. Th potential of using one swarm to modify another also depends
on how it is sensing other agents, which could be through lidar, vision, or from
a third party, such as a mother ship equipped with radar.

The sensitivity of the flocking state also has some other consequences for
modeling. For example, it shows that self-organizing real flocking biological systems which may be modeled by
swarms may have more complicated underlying dynamics and adaptive
controls than are not included in the deterministic equations of motion. There may be some other factors not modeled here which
biological flocks may use to compensate for actuation delays, noise, and
communication latency. In particular, since the linear stability of the flocking
state is neutral, there must be other individual agent controls to make the flocking more
robust to perturbations.

The preliminary results here  also show that globally coupled  swarms with delayed communication  cannot be
accurately modeled by an instantaneous model. Similar to work shown in
\cite{szwaykowska2016collective}, delays  cause new states to
emerge that are not observed in systems with instantaneous
communication. Moreover, it is also possible to create new when delay is included in
the coupling terms along with changes in network communication topology which
is not global, as shown in \cite{hindes2016hybrid}. In the case of a
given Morse potential here, simulations accurately modeling delay-coupled
swarms must take the delay into account to get accurate results. 

Finally, since only the Morse potential was used here to model interacting
swarm dynamics, other communication and coupling schemes need to be studied to
see what kinds of general statements may be made, and which conclusion pertain
only to specific swarm models. The end results of such research should prove to
open new areas of interacting and combating swarms in the future. 
\clearpage

\end{document}